%% file: main.tex
\documentclass[journal=nalefd,manuscript=article,layout=twocolumn, etalmode = truncate ,maxauthors=3]{achemso}

\usepackage{color}
\usepackage{graphicx}
\usepackage{dcolumn}
\usepackage{bm}
\usepackage{array}
\newcolumntype{L}{>{\centering\arraybackslash}m{0.33\textwidth}}
\usepackage[colorlinks=true,linkcolor=blue,citecolor=blue,urlcolor=blue]{hyperref}
\usepackage{subcaption}
\usepackage{soul}
\usepackage{amsmath}
\usepackage{amssymb}
\usepackage[dvipsnames]{xcolor}

\include{alias}

\graphicspath{{./Figs/}}

\newcommand\myaff{Laboratoire Charles Coulomb (L2C), Université de Montpellier, CNRS, Montpellier, France}

\title{
The impact of valley profile on the mobility and Kerr rotation of transition metal dichalcogenides 
}

\author{Thibault Sohier}
\affiliation{\liege}
\email{thibault.sohier@umontpellier.fr}
\alsoaffiliation{\myaff}
\author{Pedro M. M. C. de Melo}
\affiliation{\utrecht}
\alsoaffiliation{\liege}
\alsoaffiliation{\etsf}
\author{Zeila Zanolli}
\affiliation{\utrecht}
\alsoaffiliation{\etsf}
\author{Matthieu Jean Verstraete}
\affiliation{\liege}
\alsoaffiliation{\etsf}

\keywords{2D materials ; Transition metal dichalcogenides ; Electron-phonon ; Mobility ; Kerr angle ; Density-functional theory}

\begin{document}

\begin{abstract}
The transport and optical properties of semiconducting transition metal dichalcogenides around room temperature are dictated  by electron-phonon scattering mechanisms within a complex, spin-textured and multi-valley electronic landscape.
The relative positions of the valleys are critical, yet they are sensitive to external parameters and very difficult to determine directly.
We propose a first-principle model as a function valley positions to calculate carrier mobility and Kerr rotation angles.
The model brings valuable insights, as well as quantitative predictions of macroscopic properties for a wide range of carrier density.
The doping-dependant mobility displays a characteristic peak, the height depending on the position of the valleys.
The Kerr rotation signal is enhanced when same spin-valleys are aligned, and quenched when opposite spin-valleys are populated. 
We provide guidelines to optimize these quantities with respect to experimental parameters, as well as the theoretical support for \emph{in situ} characterization of the valley positions.
\end{abstract}

\maketitle

\section{Introduction}
\label{sec:intro}

Semiconducting transition-metal dichalcogenides (TMDs) hold center stage in the flatland\cite{Manzeli2017}.
They are among the most intensely studied 2D materials, along with graphene and boron nitride.
TMDs display extremely rich physics that have been explored via their optical\cite{XU2014,PhysRevLett.108.196802,PhysRevB.92.155403,PhysRevLett.119.137401,JIN2018,KIM2017}
and charge/spin transport properties\cite{Wang2012,Chu2014,Yan2019a}, most often as a function of electrostatic doping \cite{Braga2012,Gutierrez-Lezama2021,Velicky2021}.
For example, one observes: high mobility \cite{Radisavljevic2011,Radisavljevic2013,Kim2012,Lembke2012,Ovchinnikov2014,Wang2021a}; ambipolar behavior \cite{Zhang2012};insulator-metal-superconductor transitions\cite{Baugher2013,Jo2015,Costanzo2016,Piatti2018a,L.j.Li2016}; valley Hall effect\cite{Mak2014}; and a variety of bright and spin- or momentum-dark excitons\cite{PhysRevB.93.121107,PhysRevB.96.155423}.
This rich physics can be linked to a multivalley band structure and a polarized spin texture coming from strong spin-orbit interactions\cite{Xiao2012,XU2014}.
The strong and sometimes un-explained variations of macroscopic observables as a function of external parameters (strain, doping...) reflect the scattering of electrons within this complex electronic structure.
The intrinsic scattering coming from electron-phonon interactions (EPIs) is especially relevant at room temperature.

The relative position of the different valleys (at different momenta and with different spins), which we call ``valley profile'' hereafter, dictates which valleys are occupied and/or available for scattering at different doping levels or optical excitation energies.
The valley profile naturally emerges as a key parameter in the opto-electronic properties of TMDs\cite{Kormanyos2015,Yuan2016,Sohier2018,Sohier2019a,Wang2021a}. 
Our understanding of TMD physics is then challenged by the strong variations of this parameter across different TMDs, or even the same TMDs in different experimental or simulation setups.
For the latter, one notes the variability of valley profile in \ai works\cite{Kaasbjerg2012a,Kaasbjerg2013,Li2013,Jin2014,Li2015,Brumme2015,Gunst2016,Sohier2018,Sohier2019a,Gaddemane2020,Wang2021a}, highlighting a sensitivity to many factors including the exact structure, the pseudo-potentials, the exchange correlation density functional, the inclusion and type of van der Waals correction, the level of theory, etc...
This variability is physical, and also observed in experiments: 
while measurements of the conduction band profile are difficult and rare \cite{Zhang2015}, values for valence band can be found more easily \cite{Dendzik2015, Henck2018, Jin2013, Kastl2018, Kormanyos2015, Le2015, Miwa2015, Nguyen2019, Wilson2017, Yuan2016, Zhang2016a, Zhu2011}, mostly using angle-resolved photo-emission spectroscopy. 
Based on variations of the relative valley positions on the order of $100$ meV for the same material, one can infer that the valley profile depends on the environment of the 2D material and the details of the setup, including dielectric, strain, interfacial interactions with the encapsulant, or defects.
A telling example of this sensitivity to external parameters is the dependency of the valley profile on the geometric thickness of single layers of TMDs. 
A recent work \cite{Wang2021a} has shown that a variation of the thickness of $1\%$ leads to a change in the relative position of the valleys of $\approx 100$ meV.
Since the impact of thickness is very clear (in DFT and most likely experimentally) and probably dominant, it will be used in the following as a proxy for all potential factors affecting valley profile.

Gate-controlled electrostatic doping is ubiquitous in transport setups and devices based on 2D materials, and it is present in virtually all the experimental references cited here. 
The latest electrolyte gating technologies\cite{Gutierrez-Lezama2021,Velicky2021} go up to densities of a few $10^{13}$ / cm$^2$, enabling significant variations of the Fermi level (taken to be equivalent to the chemical potential throughout this work).
Similarly, the area density of excited electrons-hole pairs pumped in the material is a natural tuning parameter in optical experiments.
The quantities are different but loosely and collectively referred to as doping, and play a very important role here as they allow us to explore the valleys. 

In order to understand and control the opto-electronic properties of TMDs, it is crucial to study the impact of valley profile and doping on electron-phonon scattering.
The first consequence is on the electronic states which can be filled as a function of doping, and those that remain available for scattering. 
Another very important aspect is the impact on free-carrier screening of the EPIs. 
In addition to the usual dependency of free-carrier screening on the density of states and the shape of the Fermi surface, peculiar multi-valley screening effects have been pointed out for certain phonon modes \cite{Sohier2019a}.
The role of valley profile has been pointed out in recent experiments to explain the variation of transport measurements versus doping \cite{Piatti2018,Piatti2019,Romanin2020} or strain in TMDs\cite{Datye2022}.
This was based on an earlier theoretical study of mobility versus in-plane strain\cite{Hosseini2015}.
However, EPIs were simply modelled with deformation potentials and were assumed to be unscreened, which is incorrect for intravalley scattering, as we will see.

\begin{figure*}[htb]
  \includegraphics[width=0.4\textwidth]{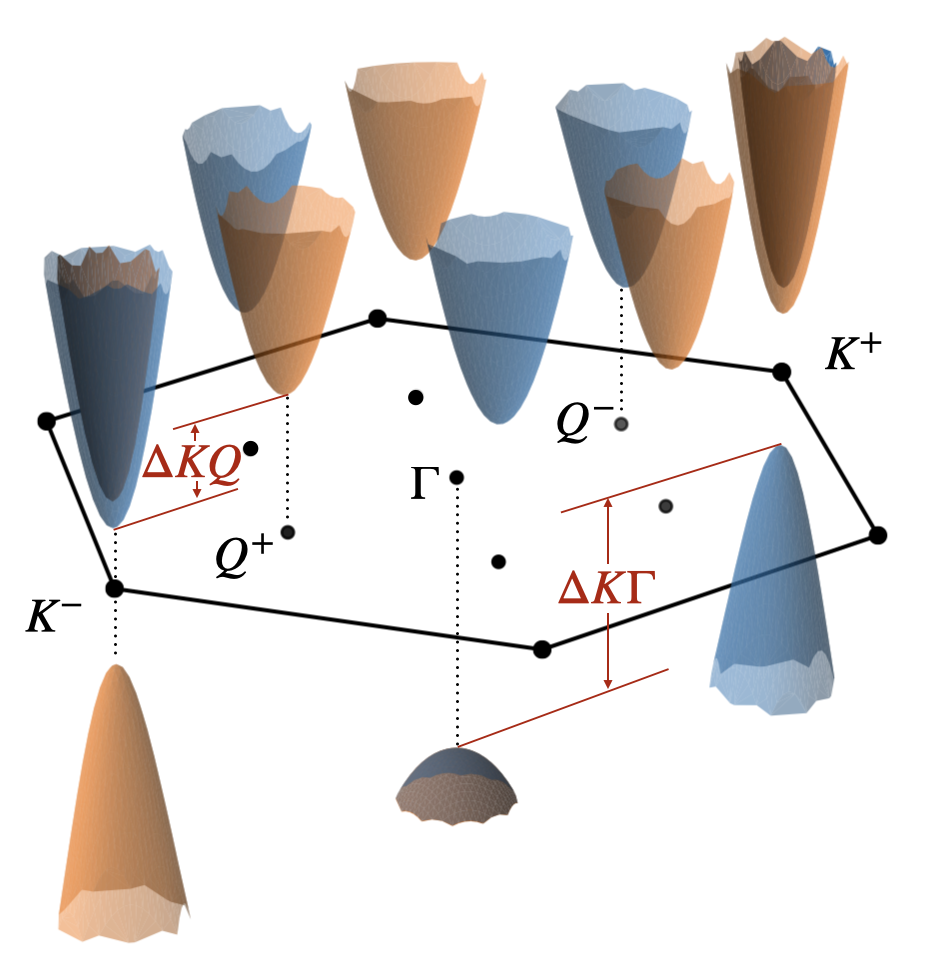}
  \includegraphics[width=0.55\textwidth]{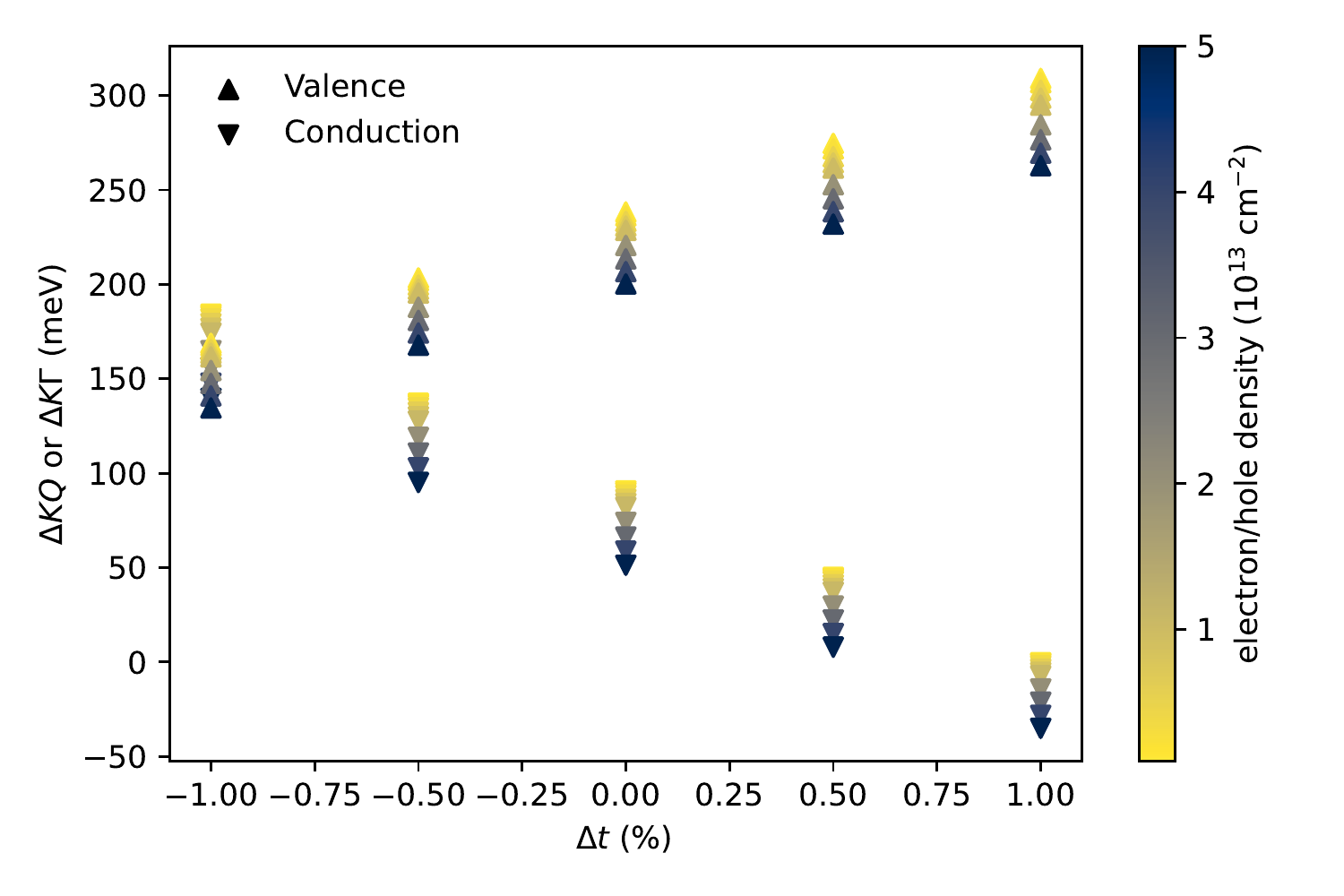}
  \caption{
(Left) Electron and hole spin-valley texture of semiconducting TMDs, for the representative case of WS$_2$. 
Orange and blue colors indicate the spin texture: up or down in the out of plane direction.
Electron valleys (top of Fig. \ref{fig:spin_val}) are situated at the corner of the hexagonal Brillouin zone (K valleys), or approximately midway between zone center and the corners (Q valleys).
Hole valleys (bottom of Fig. \ref{fig:spin_val}) are situated at the corners (K valleys) and at zone center ($\Gamma$ valleys).
Every state in the BZ is represented exactly once, such that only 2 K valleys are represented, the other 4 being equivalent by a translation following the reciprocal vectors.
Each valley is spin-split due to the spin-orbit (SO) interaction.
Note that the gap has been reduced in this visualization.
(Right) $\Delta KQ$ (left) and $\Delta K\Gamma$ (right) as a function of thickness variation $\Delta t$ and carrier density (color scale).
  }
  \label{fig:spin_val}
\end{figure*}

In this work we propose an \ai model of electron-phonon scattering to predict transport (carrier mobility) and optical properties (Kerr signal) at room temperature, with doping and valley profile as parameters.
EPIs are simulated within density functional perturbation density (DFPT).
For transport, we focus notably on modelling the impactful variation of the screening of intravalley EPIs.
Mobility is simulated within the full energy- and momentum-dependent Boltzmann equation, which is solved iteratively.
We arrive at a unified understanding of transport for both electron and holes in all semiconducting TMDs, and show how the mobility peaks as a function of doping, when the degenerate regime is reached.
When the secondary valleys are close to the band edge (low Q valley or high $\Gamma$ valley), the mobility is decreased.
We expect the transport simulations to be quantitatively accurate for clean devices at room temperature where transport is limited by phonons.
The results thus provide clear guidelines to the experimental community to optimize transport, and they suggest the possibility to reverse engineer, and characterize the valley profile from doping-dependant mobility measurements.
The Kerr signal is extracted from the solution of the Bethe-Salpeter equation\cite{10.1021/acs.nanolett.7b00175}, and we focus on the spin-population dynamics driven by intervalley EPIs, 
based on results from previous works on the impact and characteristic time scales of the EPI in temperature-dependent scattering mechanisms on TMDs\cite{MELO19,10.1021/acs.nanolett.7b00175}. 
We show that the Kerr signal is extremely sensitive to the valley energy profile of the conduction bands (while they are expected to be less important for the valence band).
This sensitivity depends on the individual TMD through differences in spin-orbit coupling strength, the thickness of the monolayer, and the different energy alignment of the valleys.
Our results show how Kerr can also be used as an indirect measurement of valley energy alignment and changes in the TMD layer thickness, by comparing the enhancement or quenching of the signal.
	
\section{Results}

The typical valley and spin structure of semiconducting TMDs is represented in Fig. \ref{fig:spin_val}, using WS$_2$ at the DFT level as an example.
The valley profile refers to the energy difference between the edges of valleys at different momenta, $\Delta KQ$ and $\Delta K\Gamma$ for the conduction and valence bands. 
Those parameters are shown in Fig. \ref{fig:spin_val}, along with their variation as a function of geometric thickness of a monolayer (for \ws, other TMDs in SI). 
Within DFT, Se compounds generally display small $\Delta KQ<100$ meV and very large $\Delta K\Gamma > 250$ meV,
while S compounds display moderate $\Delta KQ$ and $\Delta K\Gamma$ ($~ 100-200$ meV).
The thickness variations in each TMD are chosen such that $\Delta KQ$ roughly covers the  $0$ to $200$ meV range, while staying within a maximum thickness variation of $2\%$. 
Such thickness variation is experimentally accessible with pressure \cite{Nayak2015,Chen2017a}, and we expect it to be reached unintentionally in devices. 
For \ws, \wse, and \mose~this is not sufficient to bring the $\Delta K\Gamma$ under $150$ meV, making the exploration of the $\Gamma$ valley by holes unlikely. 
Spins are mostly oriented in the out of plane direction, with the relative polarities (spin up or down) of the valleys dictated by time reversal symmetry.
The spin-orbit splitting is relatively large for the Q valleys and the K valleys of the valence band, while it is weaker for the conduction K valleys, and vanishes for the $\Gamma$ valley. 
Spin splitting is overall stronger in WX$_2$ than in MoX$_2$.
The spin structure is fully accounted for, but it is not considered as a parameter, and the spin-splitting energies are kept at the DFT values.

Mobility and Kerr rotation signal are simulated as a function of valley profile and carrier density.
For transport calculations, the valley profile of both the conduction and valence band are modified.
The variations in the screening of intravalley EPIs are modelled as described in the methods section. 
In particular, we account for standard free carrier screening and propose a simple model in the context of out-of-phase multi-valley deformation potentials to account for the unusual enhancement of EPC with double occupation\cite{Sohier2019a}.
In \gw+ BSE Kerr simulations, 
given the cost of the calculations and, as noted above, the fact that the $\Gamma$ valley is in general too low to be explored by excited carriers, we focus on variations of the conduction valley profile.
We use different combinations of temperatures and chemical potentials to create populations of excited electrons and holes. 
The holes will concentrate near the K and $\Gamma$ valleys, while the electrons will be concentrated on the \kp and \qp or the \km and \qm valleys. 
The goal is to simulate an intermediate state in a pump-and-probe experiment, where the electrons have had time to scatter (conserving spin) from the \kp(\km) to the \qp(\qm) valley, but have not yet had time to scatter to valleys with opposite spin. 
No phonon contributions to the dielectric screening are taken into account, which is kept purely at the electronic level. Previous work has shown that this is enough to capture almost all essential features in the Kerr signal\cite{10.1021/acs.nanolett.7b00175,doi:10.1021/acs.nanolett.8b02774}: the main effect of the electron-phonon interaction is in the intervalley scattering, where it plays a dominant role in the process of balancing the opposite spin carrier populations. 

We now discuss the results for transport and Kerr rotation: computations and models are detailed in the Methods section below.

\begin{figure*}[h]
  \includegraphics[width=0.99\textwidth]{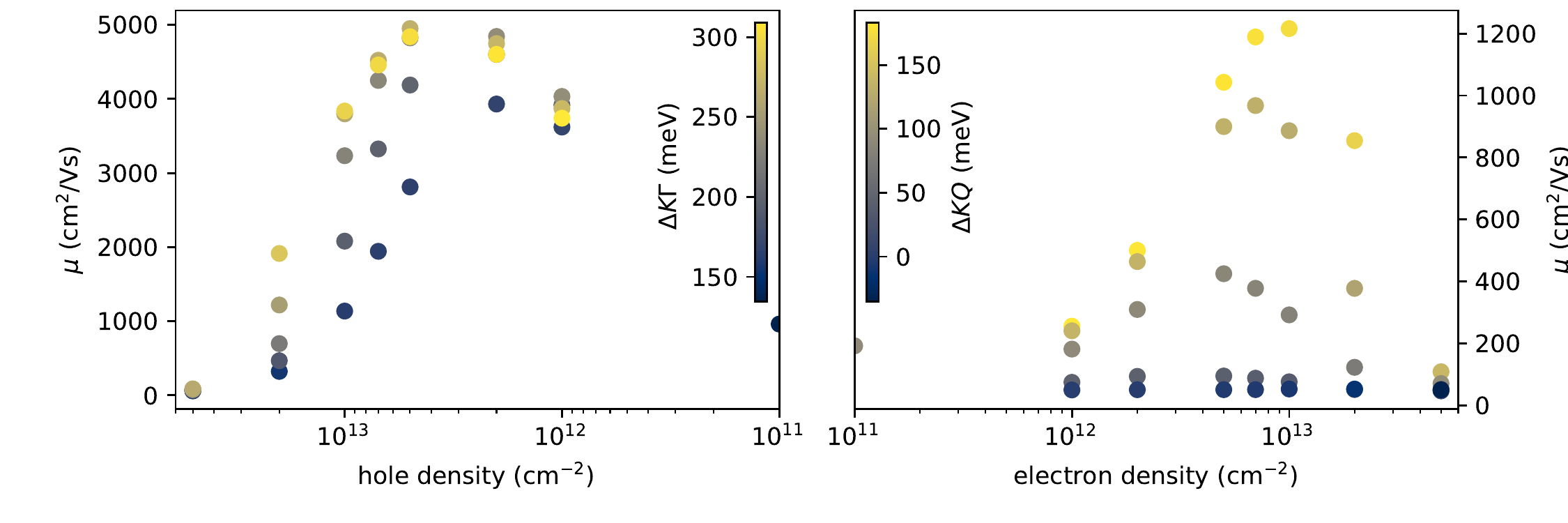}
  \includegraphics[width=0.99\textwidth]{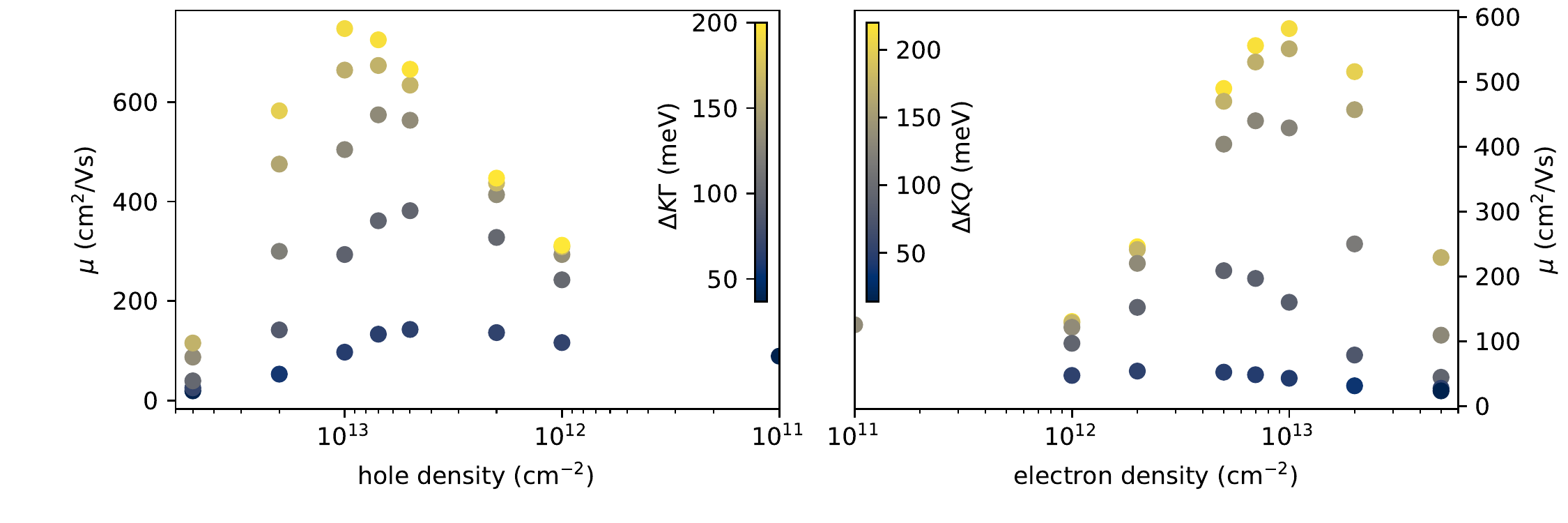}
  \caption{Room-temperature mobility as a function of electron or hole density and valley profile for WS$_2$ (top) and MoS$_2$ (bottom). In each case, the single point at $n/p = 10^{11}$ cm$^{-2}$ is the mobility of the neutral material with the relaxed structure. }
  \label{fig:mob_n_t}
\end{figure*}

\subsection{Transport as a function of electrostatic doping and valley profile}

In agreement with existing works\cite{Kaasbjerg2012a,Kaasbjerg2013,Li2013,Jin2014,Li2015,Gunst2016,Sohier2018,Sohier2019a,Wang2021a}, the phonon-limited mobility is found to be dominated by 1) intravalley scattering with TA, LA, LO, and A1 phonons, and 
2) intervalley scattering with zone boundary acoustic phonons.
Doping and valley profile generally impact scattering by modifying which states are involved.
For intra-valley scattering, an additional effect comes from free-carrier screening, with the standard doping-induced attenuation of the electron-phonon coupling for LO and TA modes, and a more peculiar multi-valley mechanism, that can actually enhance the coupling with LA and A1 modes, when K and Q (or $\Gamma$) are both occupied\cite{Frey1999,Chakraborty2012,Sohier2019a}.
Fig. \ref{fig:mob_n_t} shows the mobility of electron and hole-doped \ws\ and \mos\ as a function of carrier density for different valley profiles. 
At low doping, only the K valley is occupied and intravalley scattering dictates the behavior of the mobility. 
The  mobility increases as a function of doping, and reaches a peak around the degenerate regime, i.e. when the Fermi level hits the band edge.
This behavior was noted previously in TMDs \cite{Ma2014a,Hosseini2015}, and is expected to be general for materials dominated by screened interactions. 
The presence of a mobility peak as a function of doping points to the fact that the enhancement of the conductivity $\sigma$ due to screening is eventually overruled by the inverse dependency of mobility on carrier density, $\mu=\sigma/n$.
The position of the peak, however, is related to the details of scattering and the contribution from unscreened couplings as well.
Consequently, our findings significantly differ from previous theoretical models of doping-dependent mobility in TMDs\cite{Ma2014a,Hosseini2015}, in which EPIs models were simplified with respect to the DFPT calculations performed here.
This peak flattens when $\Delta KQ$ or $\Delta K\Gamma$ is reduced. 
Indeed, the reduction results in a larger proportion of intervalley scattering, which is associated with rather strong electron-phonon couplings which are not screened by free carriers (see SI).
In addition, multi-valley occupation causes the screening of the A1 and LA phonons to vanish (see details of the physical mechanism in the methods section).
As a result, the screening-induced peak gradually disappears, and the mobility is overall reduced.
Note that a striking electron-hole symmetry appears concerning the general behavior of mobility in \mos.
The main difference is that the spin splitting of the K valley is stronger in the valence band, leading to generally higher mobility, as the intervalley K-K' scattering involves improbable spin-flip transitions.
Otherwise, the $\Gamma$ and Q valleys play a similar role, with similar impact because the corresponding electron-phonon couplings and density of states are roughly equivalent (accounting for degeneracies).
The electron-hole symmetry is not as visible for other TMDs, because $\Delta K\Gamma$ stays relatively large within the imposed limit of $2\%$ thickness variation.
We thus arrive at a unified understanding, for electron- and hole-doped TMDs, of the role of phonon scattering in mobility trends versus doping and valley profile.

These results bring several insights and opportunities to support the experimental community. 
First, we provide the mobility for the entire range of experimentally accessible doping, when most previous \ai studies cover only the low doping limit\cite{Jin2014, Kaasbjerg2013, Li2013, Li2015}.
Given the strong variations of mobility as a function of doping, we expect this to significantly improve comparisons of theoretical and experimental data.
In addition, we use a more accurate DFPT-based model of the EPIs compared to the few previous studies considering doping dependency\cite{Ma2014a,Hosseini2015}
(see methods and SI), and quantitative comparison with experiment is meaningful.
Second, we see that a simple rule of thumb to optimize the mobility of TMDs is to work with a carrier density around $8 \times 10^{12}$ cm$^{-2}$. 
More precise values for the optimal densities can be extracted from the plots.
Third, considering the valley profile as an variable parameter enables its characterization in situ, by fitting to doping-dependent mobility measurements.
Finally, we propose the geometric thickness as an experimental knob to increase the mobility.
Unfortunately, the trend is opposite for electrons and holes: one should decrease or increase the thickness to enhance the electron or hole mobility, respectively (see Fig.~\ref{fig:spin_val}).

\subsection{Kerr as a function of excited carrier density and valley profile}

The Kerr signal measures the angle between the polarisation of incident and reflected electromagnetic fields, which is proportional to the surface magnetization density. This is often included in a pump-and-probe setup with a circularly polarized pump which creates spin populations, later measured by a linearly polarized probe. 
Due to the spin-valley coupling in TMDs, it can be used to track the populations in each valley, since the Kerr angle will only be finite if a spin imbalance exists. 
It has been shown in the past that the electron-phonon interaction plays a fundamental role in balancing the populations between valleys with different spin\cite{doi:10.1021/acs.nanolett.8b02774,MELO19,10.1021/acs.nanolett.7b00175}. 
The direct transition between \kp and \km due to EPI is forbidden, so the excited carriers start by spreading through the neighborhoods of \kp or \km in the BZ. 
States in the vicinity of \kp/\km are not pure spinors and so electrons and holes are then allowed to scatter into valleys of opposite spin. 
Nevertheless, scattering to the Q valley with the same spin polarisation is still the most favourable process, especially at higher temperatures when $\Delta KQ \sim k_B T$. 
This is indeed the process that explains the sign inversion of the Kerr signal with increasing temperature in \mose\cite{MELO19}. 

\begin{figure}[h]
    \centering
        \includegraphics[width=\columnwidth]{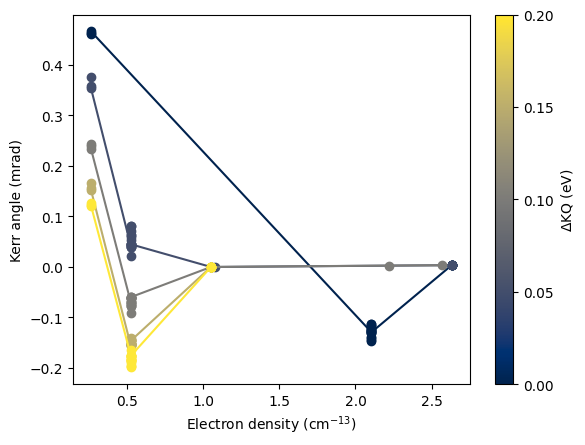}
        \includegraphics[width=\columnwidth]{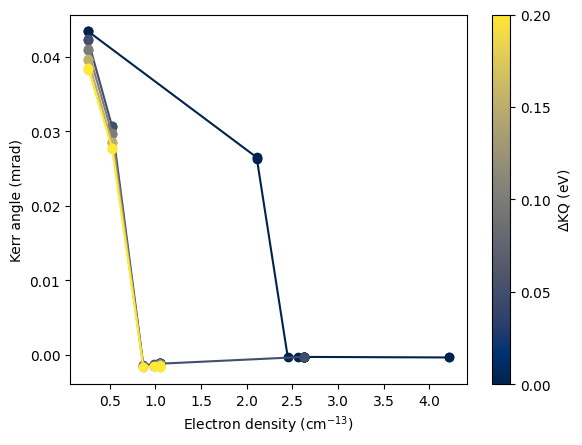}
        \caption{Amplitude of the Kerr rotation angle at the A exciton for \ws (top) and \mos (bottom) as a function of the density of excited electrons for each $\gD$KQ band alignment. The colour bar shows the level of the $\gD$KQ splitting.}

    \label{fig:kerr-2}
\end{figure}

Results are shown in Fig.~\ref{fig:kerr-2} for two materials, \ws~and \mos, with a total of 300 excited configurations per material, one for each pair $(\mu_e,T)$.
The data for \wse~and \mose~can be found in the SI.
These occupied configurations are chosen as a thermalized pseudo-equilibrium, reflecting the dominant EPI scattering channels described above.
The goal is to arrive at a qualitative relation between the Kerr rotation angle $\theta_\mathrm{Kerr}$ and the density of excited electrons for each material, depending on the splitting $\Delta$KQ. 
Once the full $\theta_\mathrm{Kerr}(\go)$ is obtained (see the SI for examples of these figures), we select the amplitude at the energy corresponding to each TMD's A exciton determined from a BSE absorption calculation, as is often done in experiments.
For the lowest value of electronic density the highest Kerr amplitude is obtained for the case where $\gD$KQ = 0: when the K and Q valleys are aligned in energy and it is easy to populate both bands with electrons. 

Since both \kp/\km and \qp/\qm valleys have parallel spin polarizations, the electron populations will reinforce the Kerr amplitude. As the $\gD$KQ energy splitting increases, it becomes harder to populate both K$^\pm$ and Q$^\pm$.
If the $\Delta KQ$ splitting is smaller than that produced by S.O.C., then it is still possible to change $(\mu_e,T_e)$ so that the lowest conduction states in $Q^\pm$ are populated.
If this is not possible, then an increase in the electronic density leads to electrons being placed in the next conduction band at K, which is strongly polarized in the opposite spin. 
In the meanwhile the density of holes must increase to enforce charge neutrality, and this contributes to reverse the net Kerr amplitude sign.
This also serves to explain why there is a significant difference in behaviour between the W and the Mo based TMDs:
The former have the largest values of energy splitting due to S.O.C., so it is easier to reach the Q valley by changing $(\mu_e,T_e)$. 
The latter have lower S.O.C.-induced energy splittings, and increasing the electronic density by changing $(\mu_e,T_e)$ will quickly place electrons in the second band in both K and Q valleys. 
This also means that for Mo-based TMDs it is easier to reach higher densities of excited electrons, especially when $\gD$KQ = 0, as it is shown at the bottom of Fig.~\ref{fig:kerr-2}.

\begin{figure}[htp]
    \centering
        \includegraphics[width=0.8\columnwidth]{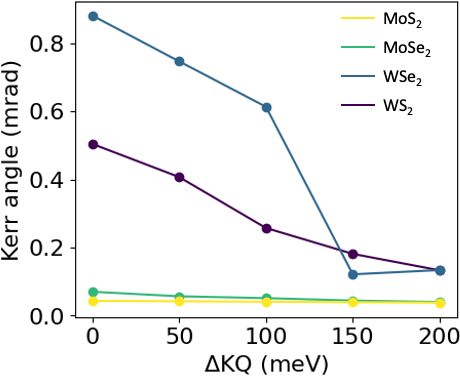}
        \caption{Change in the amplitude of the Kerr rotation for the lowest possible density of excited electrons, for each $\gD$KQ splitting (horizontal axis) and material (colour bar).}
    \label{fig:kerr-3}
\end{figure}

In Fig.~\ref{fig:kerr-3} we show how the Kerr amplitude evolves as a function of the splitting $\gD$ for each material for the lowest electron density (in the order of 10$^{11}$ cm$^{-2}$). 
Since each value of $\gD$KQ corresponds to different levels of $\gD t$, we do not represent them in the data and the colors correspond to different materials. 
Note that the correspondence between $\gD KQ$ and $\%t$ will depend heavily on the pseudo potential used during the DFT calculations. 
The data represented summarises the trend in Fig.~\ref{fig:kerr-2}: \wse{} will have the largest Kerr amplitude, followed by \ws, \mose, and \mos. 
As the $\gD$KQ splitting increases for all materials, the Kerr amplitude decreases. 
All these results show that in certain regimes the Kerr amplitude should be sensitive enough to measure the value of $\gD$KQ in a TMD, even after a change of the thickness due to the presence of a substrate or encapsulating material.

\section{Methods}
\label{sec:Methods}

\subsection{Mobility}
In this section, we describe the process to account for the main effects of valley profile and doping on EPIs and phonon-limited transport.
One could in principle simulate transport at many dopings and thicknesses, and relate the results to valley profile variations, as we have done in the past \cite{Wang2021a}.
However, too many full-system simulations would be involved, which is computationally prohibitive at this point.
A full analytical model, on the other hand, would require accounting for complex angular dependencies coming from wavefunction brackets and interference between different contributions to the electron-phonon coupling\cite{Kaasbjerg2013}.
Instead, we aim at accounting for the variations in valley profile by correcting the results of one reference \ai calculation.
We start from  high-accuracy simulations of the relaxed TMD monolayer doped at $n/p = 5 \ 10^{12}$ cm$^{-2}$ with either electrons or holes, hereafter referred to as the reference system.
We use the Quantum ESPRESSO suite \cite{Giannozzi2009,Giannozzi2017} with 2D boundary conditions and gate electrostatics \cite{Sohier2017}.
We then change the doping $n$ (or $p$) and the valley profile $\Delta KQ$ (or $\Delta K\Gamma$ ) as a post-processing step before we solve the Boltzmann transport equation (BTE) to obtain the mobility.
The BTE includes the full momentum and energy dependency of the scattering and is solved iteratively.
The general process of computing phonon-limited transport in the reference system is described extensively in \cite{Sohier2018}, with an update on the projection of the velocities (as mentioned in Ref. \cite{Sohier2020}), and the addition of spin-orbit coupling (as was already the case in Ref. \cite{Wang2021a}). 
For this work we brought a significant additional improvement that is the use of a non-uniform k-point grid to capture the Fermi surface effects more accurately. Although quite important, this is a rather technical detail that we discuss in the SI along with other computational details.

We now discuss the main methodological development for the transport part of this work: a model which includes valley profile and doping, and the associated corrections to electron-phonon scattering.
Starting from the fine band structure ($120 \times 120$ k-points grid) calculated in the neutral material,
an energy shift is applied to a given valley (e.g. Q) using our post-processing framework that groups the states into valleys\cite{Sohier2018, Sohier2020, TransportCode}. 
Solving the BTE in this new valley profile immediately takes care of the changes resulting from the energy selection rules. 
This also accounts for the impact of valley profile on the activation of intervalley scattering, as doping increases and more valleys become accessible.
Intervalley scattering is only affected by the changes in energy/momentum selection rules, the coupling matrix elements staying virtually constant.
What remains is to model the impact of valley profile on the intravalley EPI matrix elements themselves.

For the LO and TA phonon, coupling to electrons via the Fr\"ohlich and piezoelectric interactions, the matrix elements are corrected with the ratio of the dielectric functions computed as $\epsilon = 1 - v_c \chi_0$, where $v_c = \frac{2\pi}{q}$ is the Coulomb kernel in 2D. 
The non-interacting susceptibility $\chi_0$ of the added carriers, is computed similarly to our previous work\cite{Sohier2021} but in the new shifted band structure:
\begin{align}
\chi_0 &= -  \frac{2}{(2\pi)^2}  \int d^2\mathbf{k} \frac{n_{FD}(\varepsilon_{\mathbf{k}})-n_{FD}(\varepsilon_{\mathbf{k}+\mathbf{q}})}{\varepsilon_{\mathbf{k}+\mathbf{q}}-\varepsilon_{\mathbf{k}}}
\end{align}, 
where $n_{FD}$ if the Fermi-Dirac occupation and $\varepsilon_{\mathbf{k}}$ is the energy of electronic state of momentum $\mathbf{k}$. 
Band indices are summed over and omitted here.
This accounts for the increased screening of those couplings due to doping.
The A1 and LA modes, on the other hand, show multivalley screening and deformation potentials.
Counter-intuitively, free-carrier screening vanishes when doping increases beyond the point where both momentum valleys are occupied (K and Q or K and $\Gamma$).
The corresponding mechanism, described qualitatively in Ref. \cite{Sohier2019a}, is modelled in the following. We will work within the adiabatic approximation used in DFPT:
non-adiabatic effects have been discussed for the corresponding phonon dispersions \cite{Novko2020,Garcia-Goiricelaya2020}, but the consequences on electron-phonon coupling are not quantified and out of scope for the present work.
The deformation potentials associated to each valley are opposite: as the atoms are displaced according to the  A1 or LA phonons, the two types of valleys go up and down in energy in an out-of-phase fashion.
At fixed Fermi level, this implies that the charge density difference is opposite in the two valleys (one gains electrons while the other loses them).
The free-carrier response to this perturbation is proportional to the total charge difference, i.e. summed over both valleys, which implies that it vanishes as the charge density does.
We consider the isotropic response to a phonon perturbation at momentum $\mathbf(q)$, with the Fermi level and temperature as parameters.
The deformation potential $\delta V$ of each valley is the bare deformation potential $\delta V^b$ minus the potential created by the total induced density of charge $ \delta n$.
\begin{align}\label{eq:syst}
\delta V_K  &= \delta V^b_K - v_c \delta n \nonumber \\
 \delta V_Q &= \delta V^b_Q - v_c \delta n
\end{align}
Note that the "bare" potential, here, is the one without the free-carrier screening from the partially filled conduction or valence bands. 
It does include a dielectric screening that comes from the neutral material.
The dielectric screening is thus not included in the above model.
The induced density of charge has two contributions, one from the screened deformation in each valley:
$\delta n = \delta n_K + \delta n_Q =\chi^0_K \delta V_K + \chi^0_Q \delta V_Q $.
Here, $\chi^0_{K/Q}$ is the independent particle susceptibility associated to either valley, computed from the band structure as described above, but limiting the integral to states belonging to just one or the other valley.
This quantity depends on doping and temperature (chosen to be room temperature here) via the occupations, and the wave vector of the perturbation $\mathbf{q}$.
$\chi^0_{K/Q}$ tends to the valley's density of state (DOS) when $\mathbf{q} \to 0$
The bare potentials have opposite signs in each valley and are parameters of this model.

This relatively simple model explains the behavior of the DFPT EPIs very well, although it underestimates the couplings (thus potentially overestimating the mobility) for low doping ($<5\ 10^{12}$) and large $\Delta KQ$/$\Delta K\Gamma$, as  detailed in the SI.
For example, in the case where only one valley is occupied, the susceptibility of the other valley is zero, and the potential of the occupied valley is screened like a standard external perturbation.
When both valleys are occupied, the two contributions to the induced charge density are of opposite sign, reducing the overall induced potential and thus the screening as well.

The bare potentials are extracted from DFPT calculations as detailed in the SI, and the screened potentials $\delta V_{K/Q} $ are evaluated at each doping and valley profile configuration, and at each q point, by solving Eq. \ref{eq:syst}.
The EPI matrix elements of the reference calculation are then corrected as follows:
$$g_{new} = g_{ref} \times \frac{\delta V^{new}_{K, Q}}{\delta V^{ref}_{K, Q}}$$
Note that the LA mode also has a contribution from piezoelectric coupling, but it is relatively weak and quickly screened by the standard free-carrier screening.
Thus, we neglect that contribution in doped systems.
This justifies the use of a doped calculation rather than a neutral one as a reference. 
Indeed, neutral systems lead to larger piezoelectric contributions to the LA mode, making them a less appropriate starting point for the screening correction scheme where the piezoelectric contributions to LA are supposed to be absent. 
This implies a possible overestimation of the mobility towards small doping, but ensures that the main features of the mobility as a function of doping are more accurate.
All other couplings are not neglected, they are simply left as they are in the reference calculation. That is, we assume them to be doping and valley profile independent.

\subsection{Kerr}
A finite Kerr rotation angle in spin compensated systems like semiconducting TMDs implies non-equilibrium occupation of different reciprocal space valleys.
We prepare excited configurations of electrons and holes designed to reproduce the quasi-equilibrium states achieved once both systems have scattered within the initial valley where they are created, but before they have had time to travel outside that same valley. The first step is to promote electrons to the conduction bands by selecting a value for the shift in the systems chemical potential and the temperatures at which they would be, $\mu_i$ and $T_i$ with $i=e,h$, where $e(h)$ stands for electrons (holes). 
We set $T_e = T_h$, assuming that the two systems were at equilibrium. 
Energy levels are then populated according to Fermi distributions, which yield the densities of excited carriers. 
If the densities of electrons and holes do not match, meaning that the system is no longer neutral, then $\mu_h$ is tuned until such condition is obeyed. Once both densities are equal, the populations of electrons and holes in the \kp(\km) and \qp(\qm) are turned to zero, to ensure that only the other valleys have are populated with excited carriers. These are then used to evaluate the Kerr amplitude using the Yambo package\cite{Sangalli2019}.

\section{Conclusion}
We propose an \ai model of electron-phonon scattering with both doping and valley profile as parameters. 
We exploit this model to evaluate the impact of valley profile on electronic transport and Kerr rotation signal on semiconducting TMDs: MoS$_2$, MoSe$_2$, WS$_2$, WSe$_2$.
On the transport side, our findings show that all systems behave similarly for both electron and hole doping. Namely, the doping-dependant mobility peaks around $8 \ 10^{12}$ cm$^{-2}$, and the height of the peak depends on the valley profile. 
The results give quantitative guidelines to optimize the mobility, and points to the possibility of evaluating the valley profile from doping-dependant transport measurements.
Regarding the Kerr rotation amplitude, our data shows that this quantity is extremely susceptible to small changes in the valley profile, as the differences in band energies make it easier or harder to populate them. 
This suggests that Kerr amplitude measurements can be used to track the changes in valley profile of the TMD upon encapsulation, and from there infer the changes in thickness.

\begin{acknowledgement}
T.S. acknowledges support from the University of Li\`ege under the Special Funds for Research, IPD-STEMA Programme.
ZZ and PMMCM acknowledge financial support by the Netherlands Sector Plan program 2019-2023.
PMMCM and MJV acknowledge the Fonds de la Recherche Scientifique (FRS-FNRS Belgium) for PdR Grant No. T.0103.19 - ALPS, and 
ARC project DREAMS (G.A. 21/25-11) funded by Federation Wallonie Bruxelles and ULiege.
Simulation time was awarded by PRACE (project id. 2020225411) on MareNostrum at Barcelona Supercomputing Center, by
the CECI (FRS-FNRS Belgium Grant No. 2.5020.11),
as well as the Zenobe Tier-1 of the F\'ed\'eration Wallonie-Bruxelles (Walloon Region grant agreement No. 1117545).
\end{acknowledgement}

\clearpage

\section{Supporting Information}
Additional results for other TMDs; Reference transport calculations, Screening of intravalley EPIs; Kerr angle calculations.

\section{Additional results for remaining TMDs}

\begin{figure*}[htb]
  \includegraphics[width=0.49\textwidth]{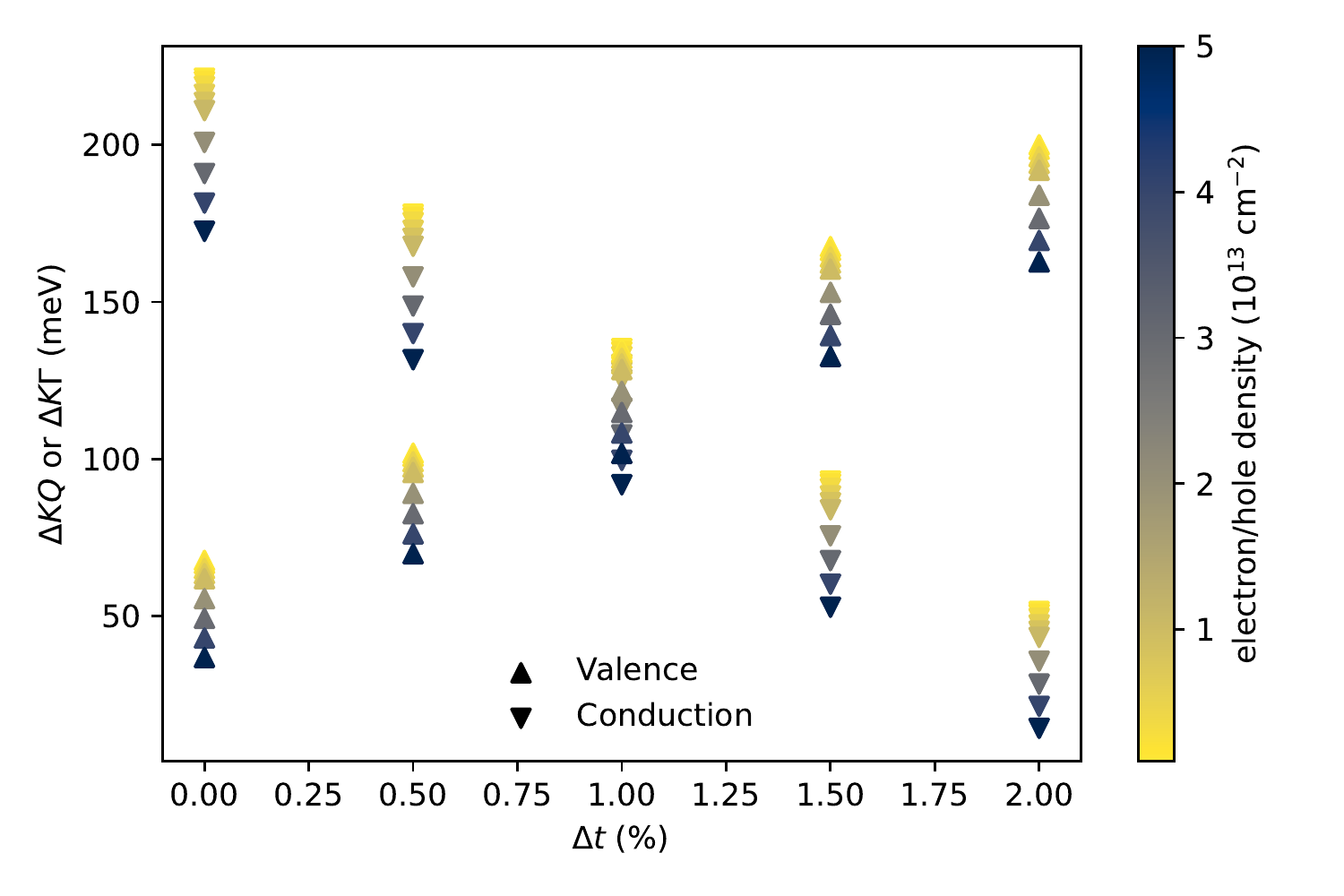}
  \includegraphics[width=0.49\textwidth]{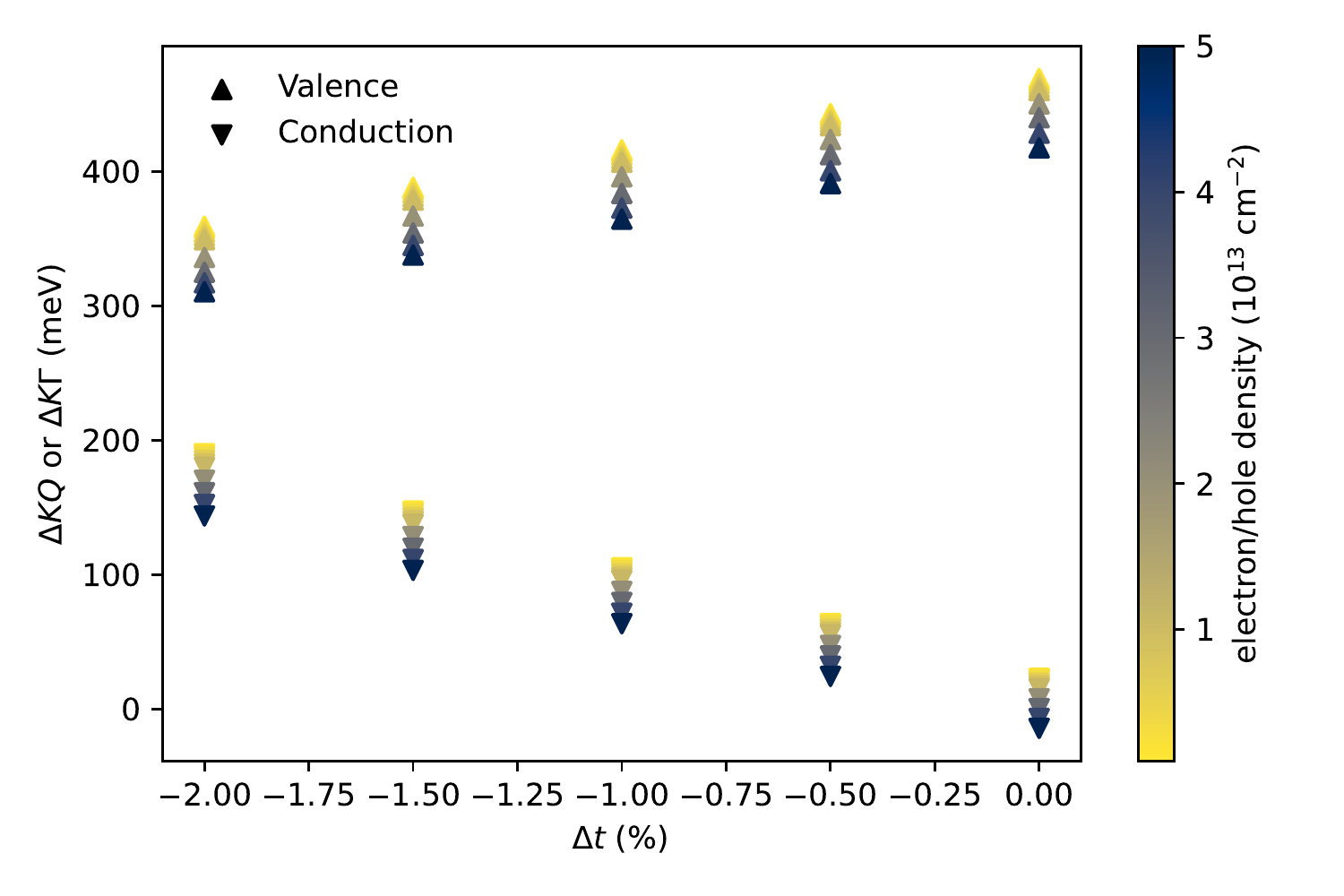}
  \includegraphics[width=0.49\textwidth]{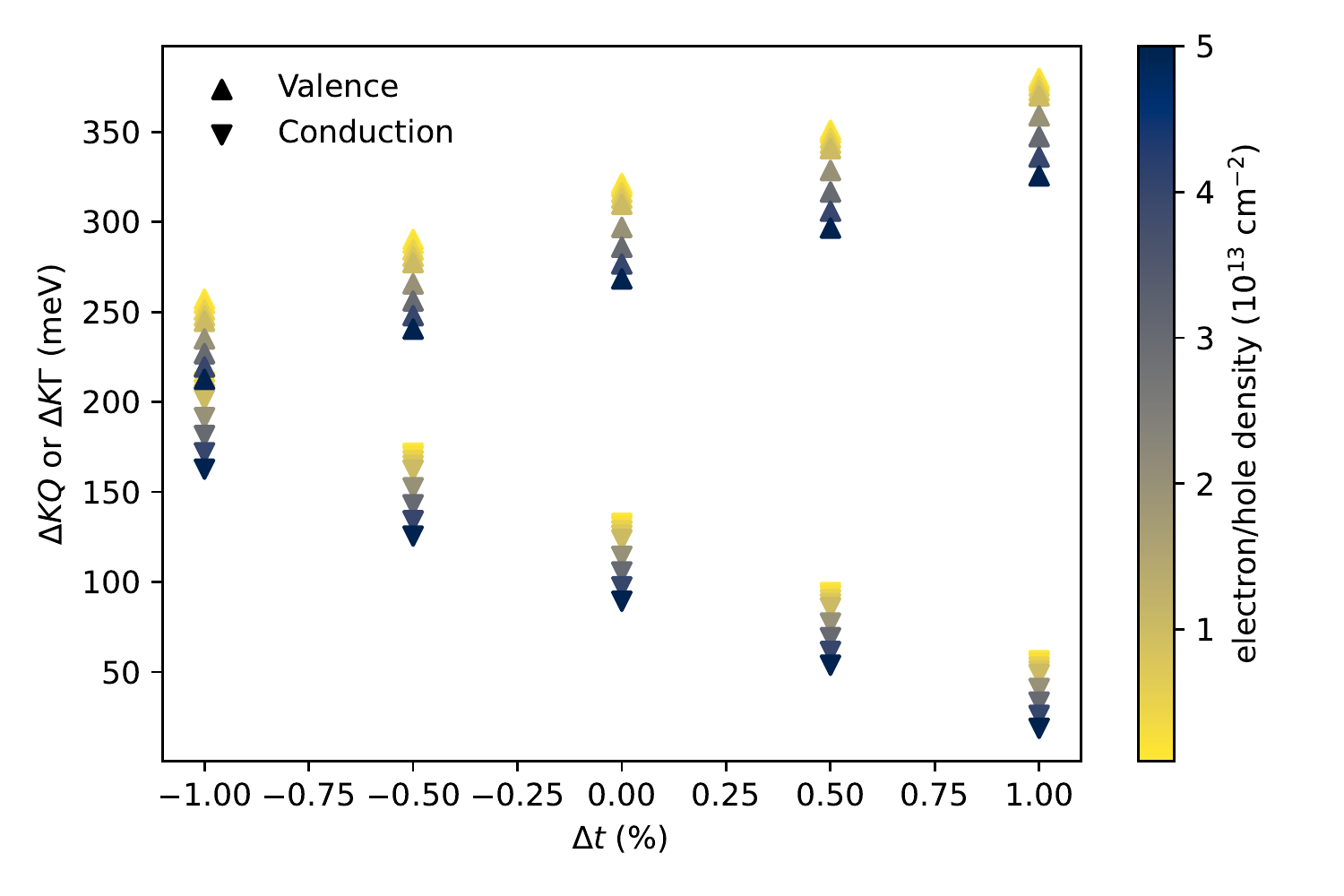}
  \caption{Valley profile as a function of doping and geometric layer thickness variation for \mos\ (left), \wse\ (right) , and \mose\ (bottom).}
  \label{fig:val_profile}
\end{figure*}

\begin{figure*}[h]
  \includegraphics[width=0.95\textwidth]{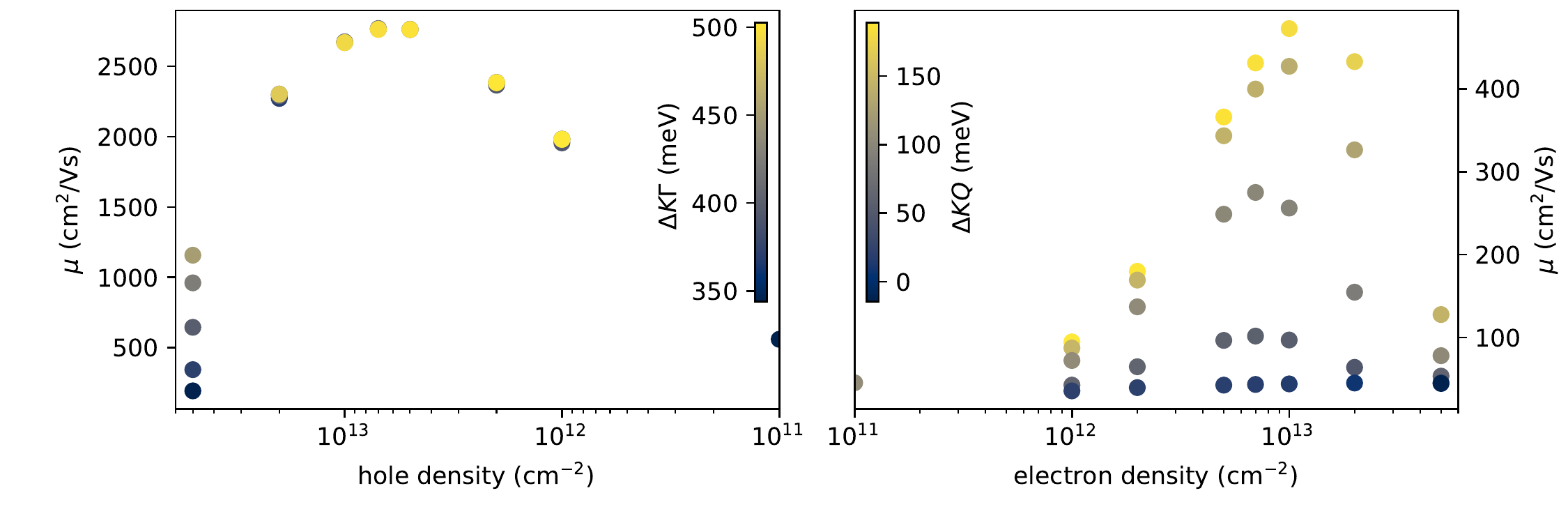}
  \includegraphics[width=0.95\textwidth]{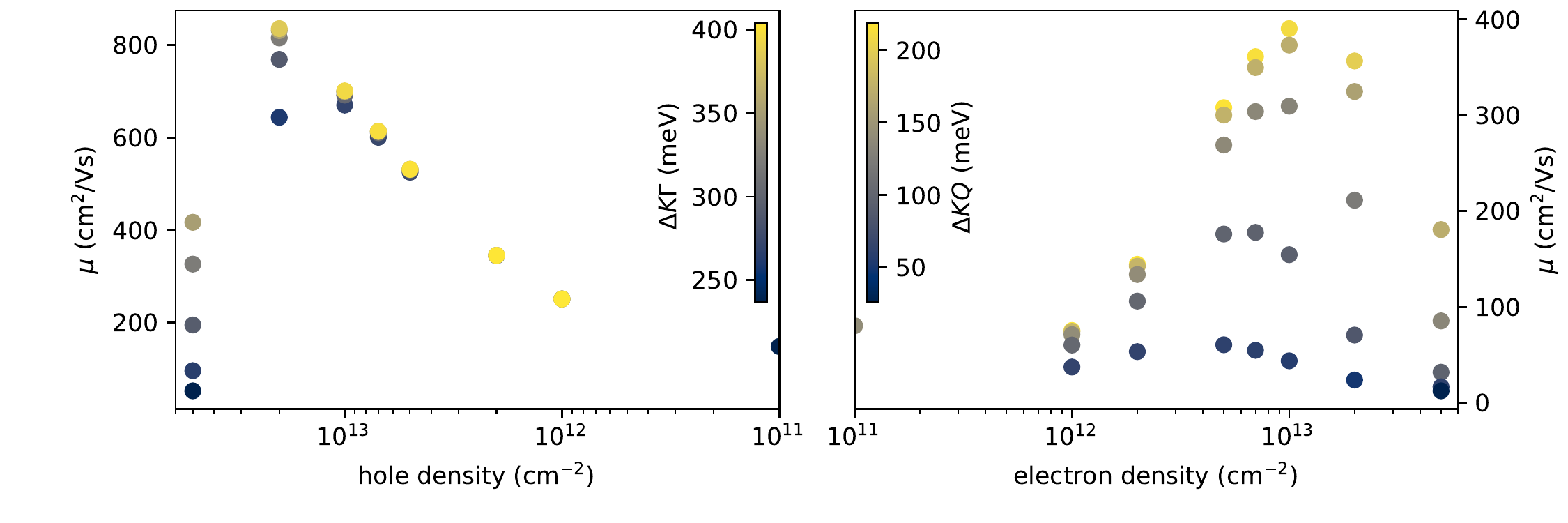}  \caption{
  Room-temperature mobility as a function of electron or hole density and valley profile for WSe$_2$ (top) and MoSe$_2$ (bottom). In each case, the single point at $n/p = 10^{11}$ cm$^{-2}$ is the mobility of the neutral material with the relaxed structure.}
  \label{fig:mob_n_t_othermats}
\end{figure*}

Fig. \ref{fig:val_profile} shows the variation of the valley profile, represented by $\Delta K Q$ and $\Delta K \Gamma$, for \mos, \wse, and \mose.
Fig. \ref{fig:mob_n_t_othermats} shows the mobility versus doping and valley profile for \wse\ and \mose.

\section{Reference transport calculations}

\begin{figure*}[h]
  \includegraphics[width=0.48\textwidth]{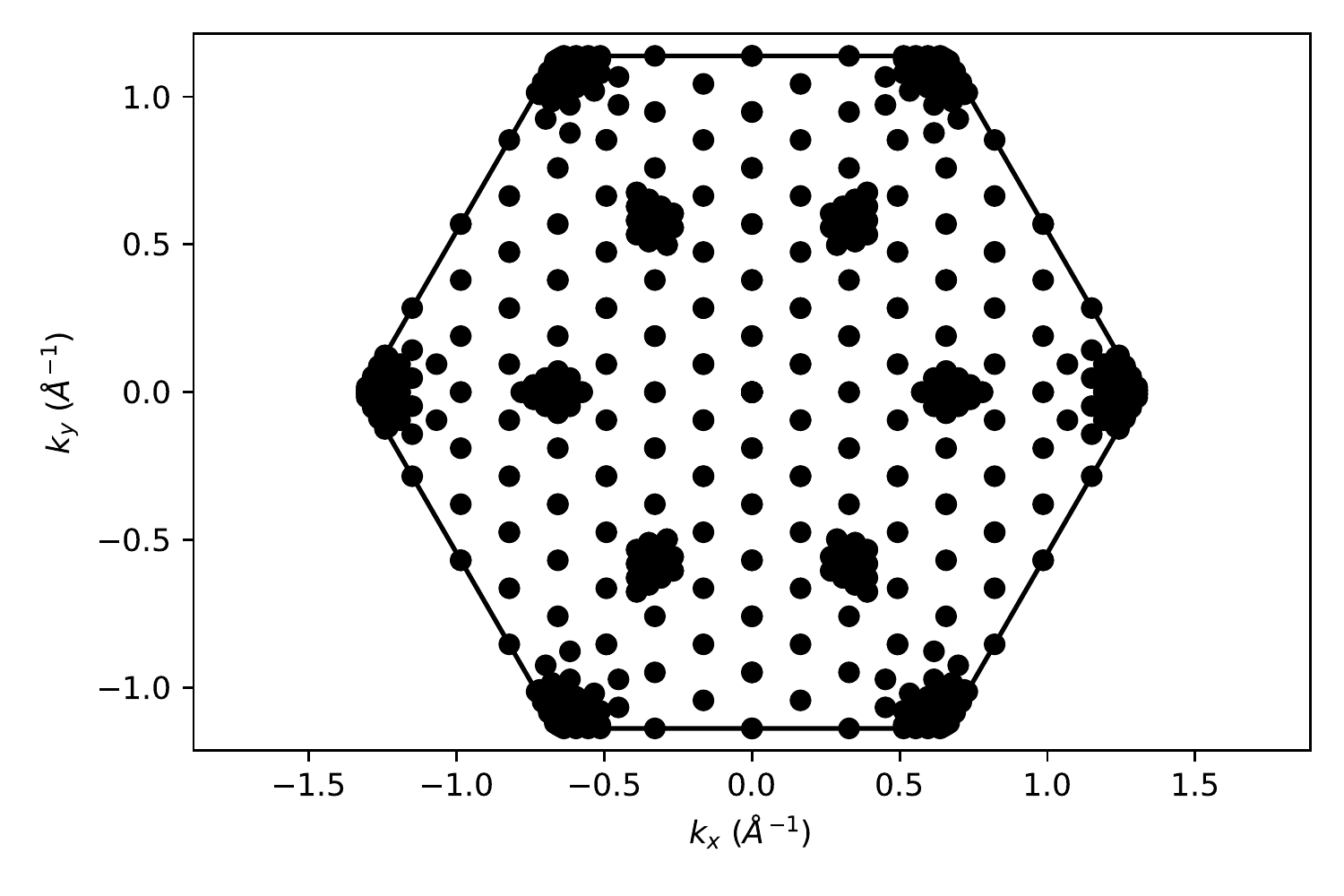}
  \includegraphics[width=0.48\textwidth]{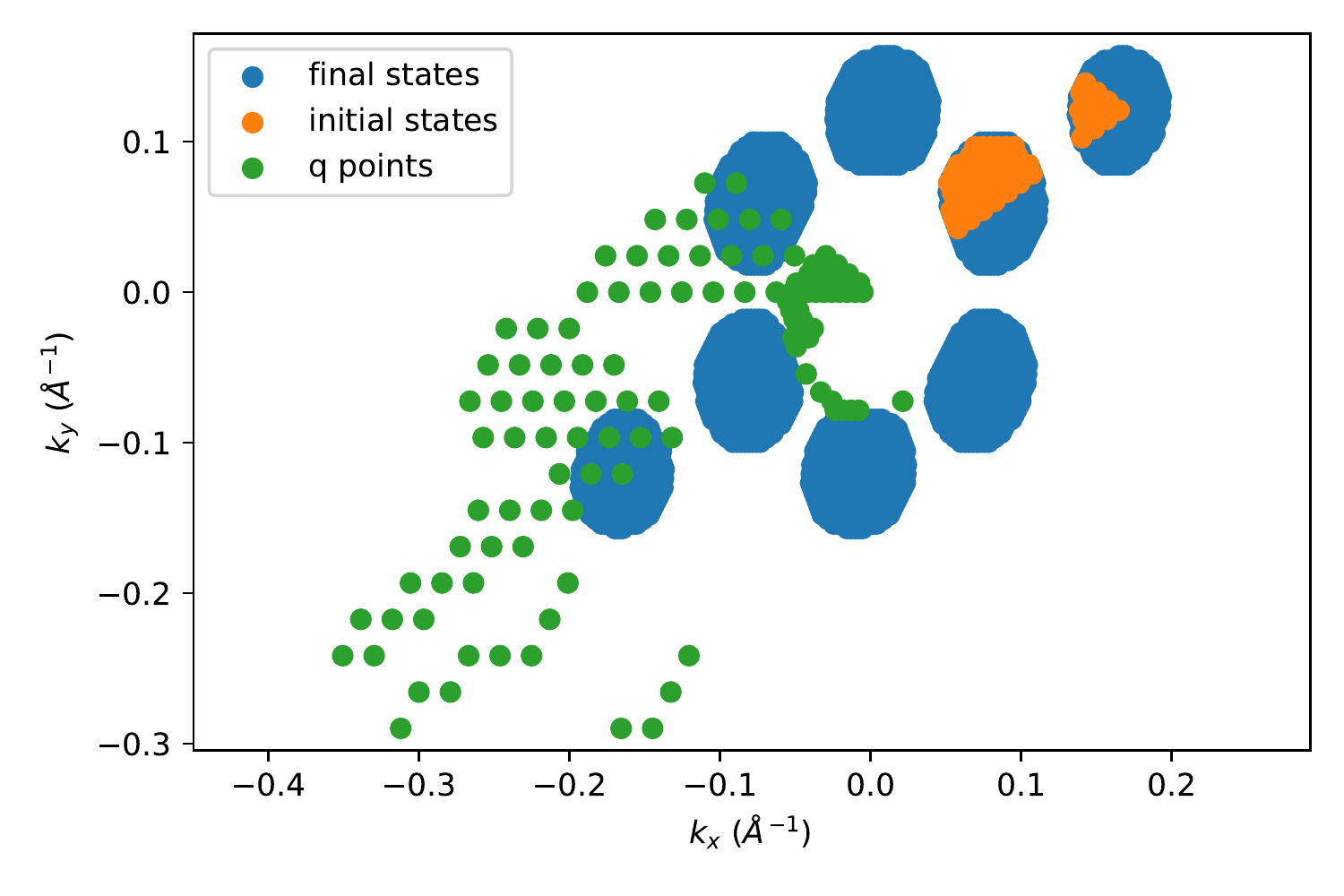}
  \caption{(Left)Sampling of k-points for DF(P)T calculations. (Right) Initial and final electronic momenta, as well as phonon momenta used for the solution to the Boltzmann equation.}
  \label{fig:sampling}
\end{figure*}

We perform full DFPT transport calculations for each TMD (n-type and p-type), at a given doping of $n/p = 5 \ 10^{12}$ cm$^{-2}$, and on the equilibrium structure (thickness relaxed).
We use norm-conserving, fully relativistic pseudopotentials\cite{Hamann2013} with Perdew-Burke-Ernzerhof (PBE)\cite{Perdew1996} functionals from the PseudoDojo library\cite{VANSETTEN201839}, with an kinetic energy cutoff of 50 (\ws), 70(\mos), or 80 (\wse, \mose) Ry.
While electrostatic doping is ubiquitous in experiments, it is not a common feature in \ai simulation.
In the Quantum ESPRESSO suite \cite{Giannozzi2009,Giannozzi2017}, the corresponding electrostatics were first developed in Ref. \cite{Brumme2015}, then adapted and included in a fully 2D framework in \cite{Sohier2017}.
As of now, this custom version of the code\footnote{\url{https://gitlab.com/tsohier/q-e/-/tree/add_gates}} is capable of computing ground state and phonon quantities in an arbitrary double gate configuration, with proper 2D boundary conditions and realistic electrostatics.

Beyond the electrostatics aspects, one needs a fine sampling of the small Fermi surface involved in doped semiconductors to account for the effects of the added free-carriers on EPIs.
This constraint is somewhat mitigated in many materials by the possibility to use a relatively large energy smearing of the occupation function - with respect to the ``real'' one which would be Fermi-Dirac at room temperature.
The standard free carrier screening is then smoothed with respect to reality, but one still captures most of the effects at a reasonable cost.
However, we have realized that it becomes problematic in materials with a complex valley profile and EPIs like TMDs, because the screening of some EPIs depends very finely on the occupations \cite{Sohier2019a}.
We thus use Fermi-Dirac energy smearing at room temperature, and tackle the sampling challenge by performing each phonon calculation with a non-uniform grid of k-points where the density is higher on the Fermi surface, as represented in Fig. \ref{fig:sampling}.
This is based on the fractal sampling proposed for graphite in Ref. \cite{Binci2021}.
The full band structure is pre-calculated on a fine grid of k-points ($120\times120$).
The valleys are filled up according to the doping of $n/p = 5 \ 10^{12}$ cm$^{-2}$ with Fermi-Dirac room temperature occupations.
We then start from a typical $12 \times 12$ Monkhorst-Pack k-points grid.
For each parallelogram in the grid, if an occupied conduction band state is found within, we tesselate it further, otherwise it is left as is.
The process is repeated 3 times, the grid density quadrupling each time.
We thus end up, locally on the Fermi surface, with the equivalent of a $96 \times 96$ grid.

The workflow to solve the BTE, detailed in Ref. \cite{Sohier2018}, involves 3 separate momentum samplings for initial and final states as well as the phonons involved in all the relevant transitions. The sampling of the final states is the finest: a $120 \times 120$ Monkhorst-Pack k-points grid on which we compute the band structure (as a non self-consistent step after a typical ground state calculation in the neutral material) and velocities.
The initial state are sampled on a grid twice as coarse.
Finally, the phonon momenta are obtained by linking all initial states with a portion of the final states (half for intravalley, an eighth for intervalley) and reducing the result by symmetry.
This usually yields around 100 to 200 irreducible phonon momenta.

\begin{figure*}[h]
  \includegraphics[width=\linewidth]{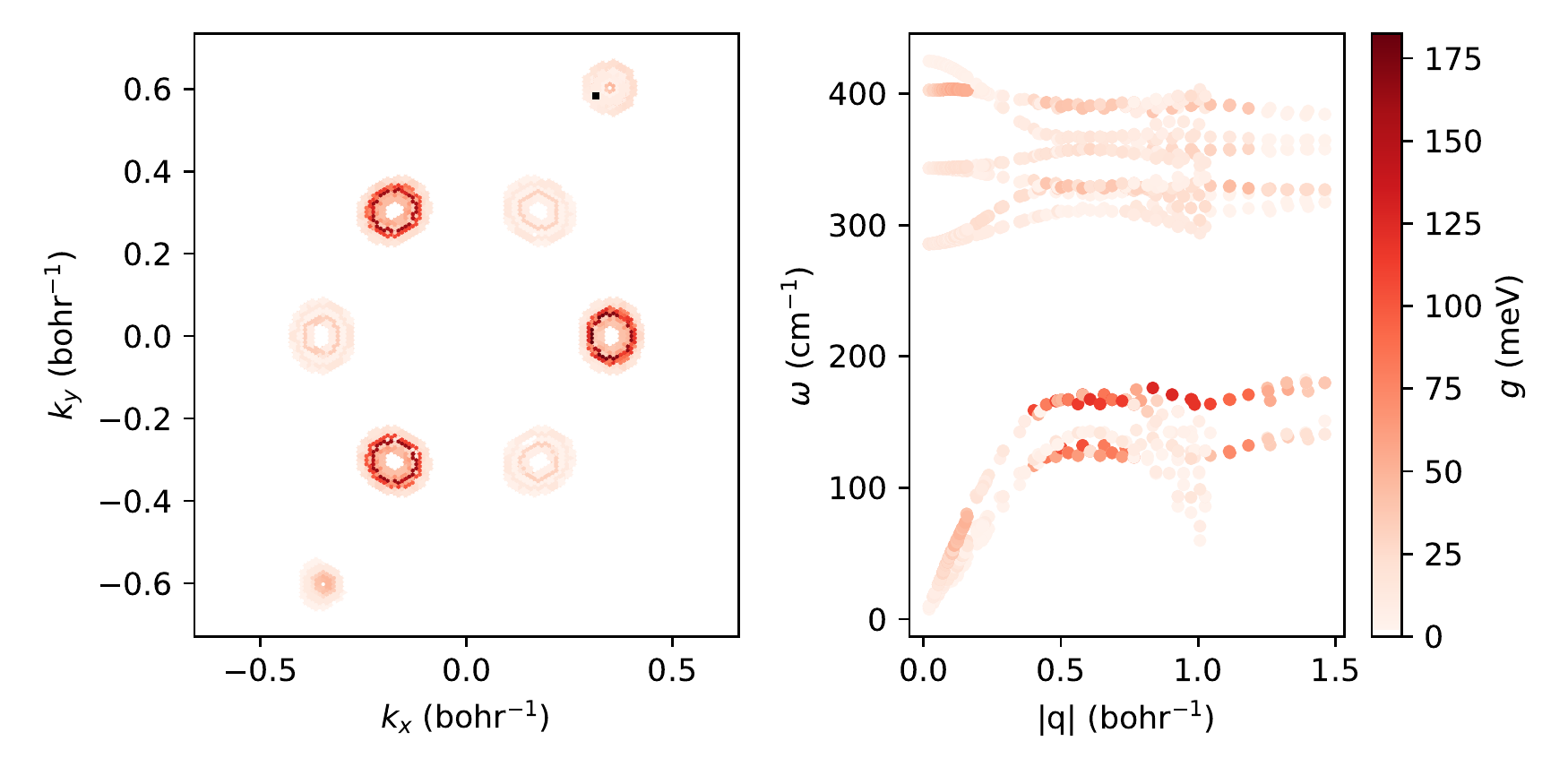}
  \caption{Electron phonon coupling matrix elements (color scale) for \ws\ as computed in the reference calculation. In the left panel, we select an initial state in the K valley (located with a black square), apply the energy and momentum selection rules, multiply by the occupation of the phonon modes at room temperature, and sum over all modes to get a total, effective EPC scattering strength. On the right, we simply plot the EPC matrix element for each momentum and mode. }
  \label{fig:EPC_map}
\end{figure*}

Electron-phonon coupling matrix elements are computed for each relevant pair of initial and final state. They are represented in Fig. \ref{fig:EPC_map}. Clearly, the phonons with the highest couplings, by far, are zone border acoustic phonons. They correspond to intervalley (K-Q) transitions. Thus, when this intervalley scattering is activated according to energy selection rules, it dominates transport. The next most significant couplings come from the LA (highest acoustic branch) and A1 (second highest optical branch) modes around $q=0$. 

\section{Screening of intravalley EPIs}

\begin{figure*}[h]
  \includegraphics[width=\linewidth]{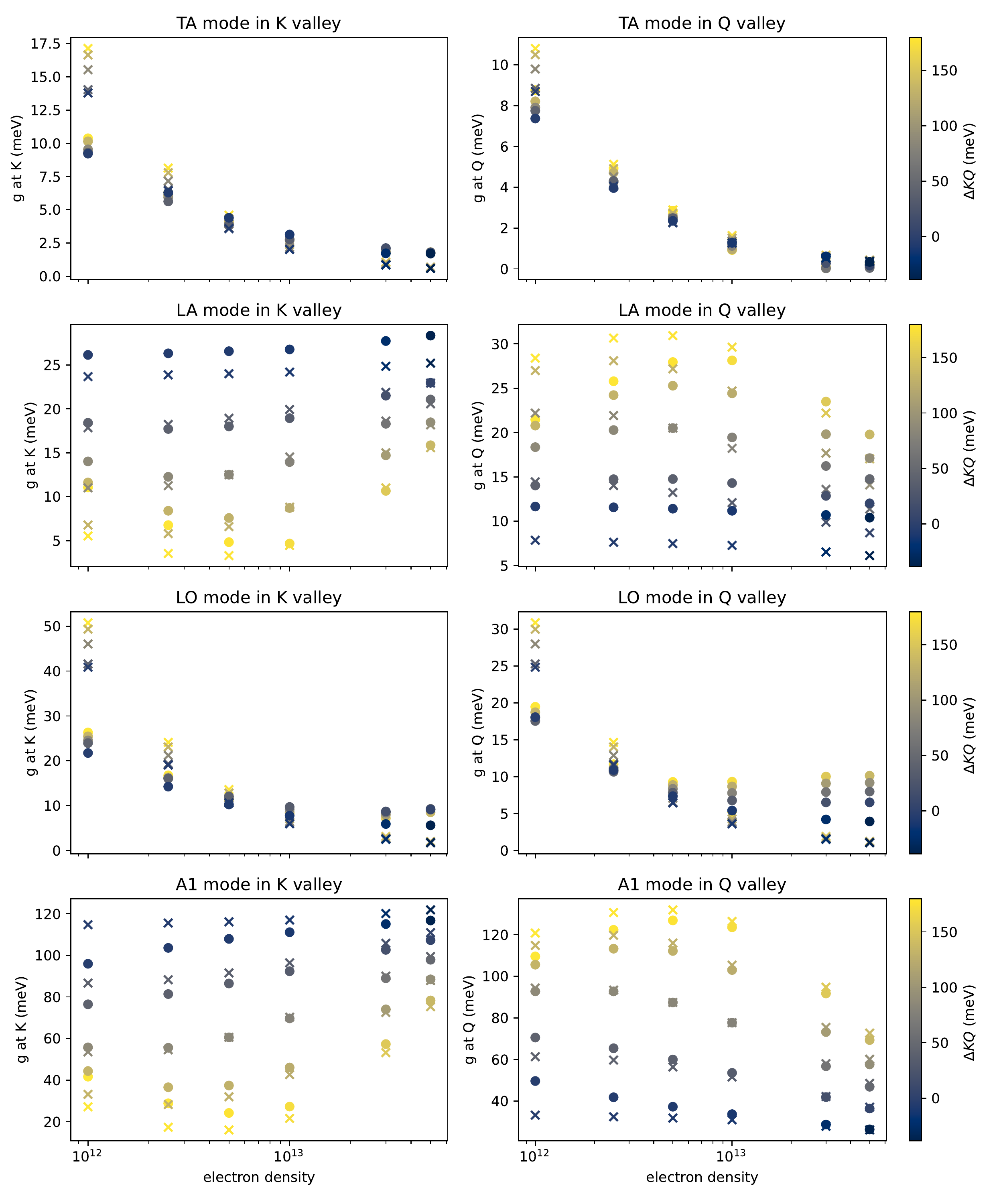}
  \caption{Electron phonon coupling for the main phonon modes in WS2 versus doping $n$ and valley profile $\Delta KQ$. We pick either K (left) or Q (right) as an initial state, and a small q point in the $\Gamma-K$ directions. Dots are direct DFPT calculations, crosses are the model.}
  \label{fig:EPC_n_t}
\end{figure*}

Here, the model for the effect of doping and valley profile on the free carrier screening of intravalley couplings is discussed and detailed.
First, we perform some relatively simple DFPT calculations at different electron doping $n$ and thicknesses $t$.
The variation in thickness is associated with the  $\Delta KQ$ variations shown in Fig. \ref{fig:val_profile}.
The ground state calculation is done on a $32 \times 32$ k-point grid and Marzari-Vanderbilt smearing at $0.01$ Ry.
We are not interested in the realistic nature of the screening effects, here, we simply wish to demonstrate their behavior as a function of doping and valley profile.
Fig. \ref{fig:EPC_n_t} shows the EPI matrix element at a given q point (small $|q|=0.023$ \AA$^{-1}$, in the $\Gamma-K$ direction) as a function of doping and valley profile (via thickness), for TA, LA, LO and A1g.
Direct DFPT (dots) is compared to the model (crosses), with all the screening quantities computed with the same bands, k-points sampling and smearing as the DFPT calculations.
In the $\Gamma-K$ direction, the piezoelectric coupling is maximal for TA and it vanishes for LA\cite{Kaasbjerg2013}, so that we can separate the two types of couplings.

The expected qualitative behaviors of the couplings are confirmed.
The TA and LO phonons couple via the piezoelectric and Fröhlich interaction, respectively.
Being mediated by electric fields, those are screened by the free-carriers in the standard way: as $n/p$ increases the coupling is killed rather rapidly.
For those couplings, most of the discrepancy between model and DFPT can be explained by the fact that we assume the entire coupling to be field-mediated, while in reality there is also a small non-screened contribution. 
The odd DFPT results for the LO mode at high doping (especially in the Q valley), are due to a mix of the LO and TO mode, the TO mode having a fairly constant coupling that is stronger in the Q valley.


The behavior of the couplings to the LA and A1 phonons is quite different.
In particular, it strongly depends on the valley profile and the occupation of the valleys.
Small doping and large $\Delta KQ$ corresponds to the case where only the K valley is occupied.
In that situation, we see the doping-dependency of the coupling around K is similar to the one induced by standard free carrier screening.
As the doping increases, however, the coupling re-increases as the electrons start to feel the presence the Q valley.
It is interesting to note that the coupling at Q follows an opposite trend, first increasing and then decreasing.
The other extreme case is that of vanishing $\Delta KQ$, when both K and Q valley are quickly occupied even at small doping.
In that situation the couplings are virtually doping independent, indicating the absence of free-carrier screening.
As the valley profile varies between those two cases, intermediary behaviors are observed.
An important characteristic that one can deduce from those curves is that the behaviors of the coupling at K and Q are very much tied to one another. 
In fact, the sum of the two is almost constant (variations below $10\%$) as a function of doping and valley profile.
Thus, when the coupling of the occupied valley is screened by free carriers, the coupling in the unoccupied valley is actually enhanced.
To the best of our knowledge, this behavior has not yet been described.
While the enhanced coupling in the empty valley is not relevant for transport, it could play a role for the dynamics of excited carriers in doped TMDs.
Concerning the comparison with DFPT for LA and A1 phonons, the model reproduces the calculations quite well in general given the complexity of the mechanisms involved. 
The discrepancies most likely come from the relative simplicity of the model.

All the discrepancies seen in this figure would lead to the largest error when those intravalley couplings dominate the transport, i.e. at small doping and for large $\Delta KQ$, when only the K valley would be occupied.
Looking a the couplings in the K valley in this regime, we see that a fortuitous cancellation of errors might occur between an overestimation of the coupling for TA/LO and an underestimation for LA/A1.
Thus, overall, we don't expect a significant error on the mobility from those discrepancies.


\subsection{Evaluation of bare potentials}

The absence of doping dependency of the EPIs when both valleys are occupied in the DFPT calculations of Fig. \ref{fig:EPC_n_t} indicates that free-carrier screening vanishes, and thus, $\chi^0_K \delta V_K + \chi^0_Q \delta V_Q = 0$ and $\delta V_{K/Q} = \delta V^b_{K/Q}$
Thus, $\frac{\delta V^b_K}{\delta V^b_Q} \propto \frac{\chi^0_Q}{\chi^0_K} \propto \frac{DOS_Q}{DOS_K}$.
The last relationship is valid in the $\mathbf{q} \to 0$ limit, but we assume it can be extended at finite momenta.
Finally, in agreement with the \ai results above, we assume that the quantity $\delta V_K - \delta V_Q =  \delta V^b_K - \delta V^b_Q = \Delta V^b $ is a constant (independent of doping and valley profile) determined solely by the electron-phonon coupling strength. 
$\Delta V^b $ can show a slight dependency on the momentum, especially for the coupling with acoustic phonons. 
In such cases, we use the value extracted around $q\approx 0.1$ \AA$^{-1}$, which correspond to a rough estimation of the size of the Fermi surface at $\approx 50$ meV above the bottom of the K valley. 
Using those remarks the bare potentials can be evaluated from the $\Delta V^b$ and the DOS of the valleys as:
\begin{align}
V^b_K &= \frac{\Delta V^b}{1+\frac{DOS_K}{DOS_Q}} \\
V^b_Q &= \frac{\Delta V^b}{1+\frac{DOS_Q}{DOS_K}}
\end{align}
This is a convenient procedure since it involves quantities that do not depend on doping and valley profile.



\section{Kerr}

\begin{figure}
    \centering
        \includegraphics[width=\columnwidth]{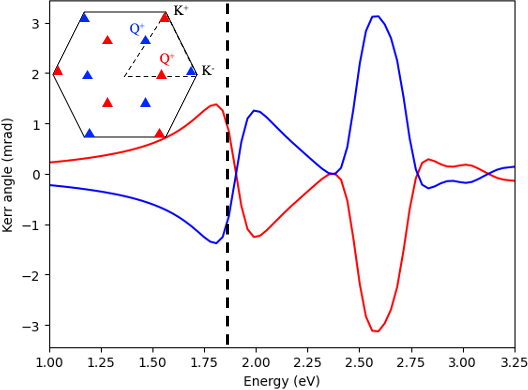}
        \includegraphics[width=\columnwidth]{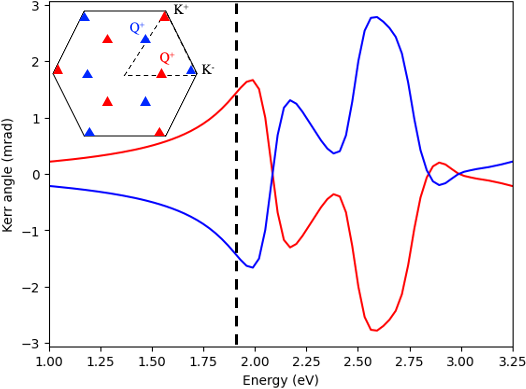}
        \caption{Amplitude of the Kerr rotation angle at the A exciton for \ws~(top) and \wse~(bottom) from the \qp (red) and \qm (blue) configurations for T = 40 K, $\mu_e$ = 0 eV, and $\gD$KQ = 0 eV. The symmetry of the signal reflects the opposite spin polarisation of each valley occupation configuration. The inset shows the regions of the Brillouin zone where electrons are promoted to the conduction bands in both the \qp and \qm configurations. The black dashed vertical line marks the energy of the A exciton for each system.}
    \label{fig:kerr-4}
\end{figure}

The geometrical and ground state electronic properties of \ws{}, \wse{}, \mose{}, and \mos~were computed with DFT. 
We employed 24x24x1 k-point grids, with 120 Ry kinetic cutoff, using PBE-GGA full relativistic pseudo potentials from PSEUDO-DOJO\cite{VANSETTEN201839}. 
To remove the interaction between replica images we used 30 Bohr of vacuum between each layer. 
Geometry optimization was performed until diagonal stress components and forces were smaller than 0.01 kBar and  $10^{-6}$ Ry/Bohr, respectively. 
Electronic band structures were then computed on a 12x12x1 k-point grid. 
The energy levels of the ground state conduction bands were deformed in order to reproduce $\gD$KQ = 0, 50, 100, 150, and 200 meV energy splittings. 
Full frequency \gw calculations were performed to obtain the quasi-particle band energies.
Realistic excited state configurations were then created by removing all symmetries from system and populating the bands using Fermi distributions with different values of ($\mu_e, T_e;\mu_h, T_h$), with the restriction that the final densities of electrons and holes were equal. 
We then created two excited state configurations for each ($\mu_e, T_e;\mu_h, T_h$) quartet, with excited electrons and holes populating either the \kp and \qp or the \km and \qm valleys shown in Fig.~\ref{fig:spin_val}. T
hese new configurations were then used to calculate the correction to the energy levels induced by the presence of excited carriers. 
With it we can obtain the full correction to the energy levels given by\cite{PhysRevB.84.245110,PhysRevB.93.155102,10.1021/acs.nanolett.7b00175}
\begin{multline}
    \varepsilon_{\nk}^\mathrm{NEQ} = \varepsilon_{\nk}^\mathrm{DFT} +\gD\varepsilon_{\nk}^\mathrm{G_0W_0-dyn} + \\
    +\gD\varepsilon_{\nk}^\mathrm{COHSEX}[\gr_e = 0] - \gD\varepsilon_{\nk}^\mathrm{COHSEX}[\gr_e \neq 0],
    \label{eq:e-corr}
\end{multline}
where $\varepsilon_{\nk}^\mathrm{DFT}$ are the DFT energy levels; $\gD\varepsilon_{\nk}^\mathrm{G_0W_0-dyn}$ the dynamical \gw corrections; $\gD\varepsilon_{\nk}^\mathrm{COHSEX}[\gr_e = 0]$ the \gw COHSEX corrections with the system at equilibrium; $\gD\varepsilon_{\nk}^\mathrm{COHSEX}[\gr_e \neq 0]$ the \gw COHSEX correction for when the system has a finite density of excited electrons and holes. In both dynamical and COHSEX \gw we used the kinetic cutoffs 20 Ha, 5 Ha, and 4 Ha for the G-vectors in the FFT, exchange self-energy, and correlation self-energy. 
A total of 300 bands were needed to converge the electronic screening.
The last step is the calculation of the Kerr amplitude, which involves computing the dielectric matrix $\epsilon_{\ga\gb}(\go)$ with the Bethe-Salpeter equation (BSE) to accurately include the presence of excitonic states. 
This involves diagonalizing the BSE Hamiltonian
\begin{multline}
    H_{\substack{cv\kk \\ c'v'\kk}} = ( \varepsilon_{c\kk}^\mathrm{NEQ} - \varepsilon_{v\kk}^\mathrm{NEQ})\gd_{cc'}\gd_{vv'} +\\
    + \mathrm i[f_{c\kk}(T_e) - f_{v\kk}(T_h)]\left(W_{\substack{cv\kk \\c'v'\kk}} - 2V_{\substack{cv\kk \\c'v'\kk}}\right),
    \label{eq:bse-h}
\end{multline}
where $W$ and $V$ are the screened direct and exchange interaction, respectively. 
The kinetic cutoffs and empty bands used were the same as the \gw calculations, and a total of six bands were used in the kernel to set up the electron-hole transitions. Once the full set of exciton energies and eigenvectors is obtained, ${E_\lambda, A_{cv\kk}^\lambda}$, we can build the dielectric tensor using
\begin{multline}
    \epsilon_{\ga\gb}(\go) = \delta_{\ga\gb} - \\
    \sum_{\ga\gb}\sum_{\substack{cv\kk\\c'v'\kk'}} d_{cv\kk}^\ga\left(d_{c'v'\kk'}^\gb\right)^*\frac{A_{cv\kk}\(A_{c'v'\kk'}\)^*}{\omega - E_\lambda + i0^+},
    \label{eq:eps}
\end{multline}
where $d_{cv\kk}^\ga$ are the dipole matrix elements, and $\ga,\gb = x,y,z$. The Kerr amplitude can then be evaluated using
\begin{equation}
    \theta_\mathrm{Kerr}(\go) = -\Im\left[\frac{\epsilon_{xy}(\go)}{\epsilon_{xx}(\go)\sqrt{\epsilon_{xx}(\go) + 1}}\right].
    \label{eq:kerr}
\end{equation}
Results for the Kerr angle are shown in Fig.~\ref{fig:kerr-4} for \ws{} and \wse{} for T = 40 K, $\gd\mu_e$ = 0 eV, and $\gD$KQ = 0 eV. 
The red/blue line shows the $\theta_{\mathrm {Kerr}}(\go)$ dispersion when the the \qp/\qm valleys are occupied. 
The symmetry of the vallues of $\theta_{\mathrm {Kerr}}(\go)$ is an important check, as it proves that our method does indeed respect the underlying Physics of each system. 

\begin{figure}
    \centering
        \includegraphics[width=\columnwidth]{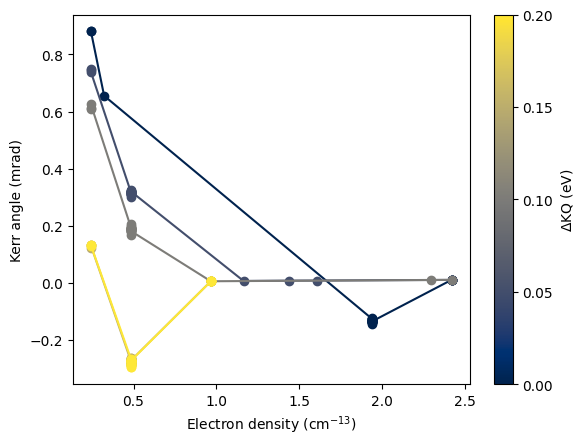}
        \includegraphics[width=\columnwidth]{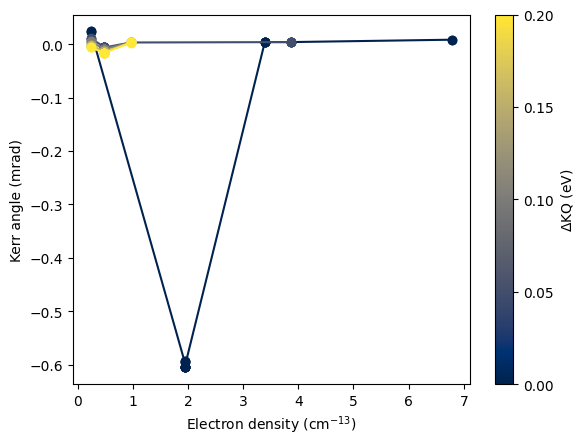}
        \caption{Amplitude of the Kerr rotation angle at the A exciton for \wse{} (top) and \mose{} (bottom) as a function of the density of excited electrons for each $\gD$KQ band alignment. The colour bar shows the level of the $\gD$KQ splitting.}
    \label{fig:kerr-5}
\end{figure}

\clearpage

\bibliography{QV}

\end{document}

%% file: alias.tex
\newcommand{\be}{\begin{equation}}
\newcommand{\ee}{\end{equation}}
\newcommand{\bea}{\begin{eqnarray}}
\newcommand{\eea}{\end{eqnarray}}

\newcommand{\ws}{WS$_2${}}
\newcommand{\mos}{MoS$_2${}}
\newcommand{\wse}{WSe$_2${}}
\newcommand{\mose}{MoSe$_2${}}

\def\kk		{{\bf k}}

\def\ga         {\alpha}
\def\gb         {\beta}

\def\gd         {\delta}
\def\gD         {\Delta}

\def\go         {\omega}

\def\gr         {\rho}

\renewcommand{\)}{\right)}

\def\gw      {G$_0$W$_0$~}
\def\mose {MoSe$_2$}

\def\nk         {{n{\bf k}}}

\def\kp          {K$^+$}
\def\km          {K$^-$}
\def\qp          {Q$^+$}
\def\qm          {Q$^-$}

\def\bverb      {\renewcommand{\baselinestretch}{1}\begin{verbatim}}
\def\everb  	{\end{verbatim}\renewcommand{\baselinestretch}{1.3}}

\def\ai         {{\it ab initio}~}

\def\ga         {\alpha}
\def\gb         {\beta}
 
\def\gD         {\Delta}

\def\go         {\omega}

\def\kk         {{\mathbf k}}

\def\nk         {n\kk}

\renewcommand{\)}{\right)}

\newcommand{\etsf} {European Theoretical Spectroscopy Facility (ETSF) www.etsf.eu}
\newcommand{\liege}{nanomat/Q-mat/CESAM and Department of Physics, Universit\'e de Li\`ege, B-4000 Sart Tilman, Li\`ege, Belgium}

\newcommand{\utrecht}{Chemistry Department, Debye Institute for Nanomaterials Science, Condensed Matter and Interfaces, Utrecht University, PO Box 80.000, 3508 TA Utrecht, The Netherlands}